\documentclass[a4paper,11pt]{elsarticle}
\usepackage{lineno,hyperref}
\hypersetup{colorlinks=true}

\usepackage[utf8]{inputenc}
\usepackage[official]{eurosym}

\usepackage{amsmath}
\usepackage{IEEEtrantools}
\usepackage{mathtools}
\usepackage{siunitx}
\usepackage{tabularx}
\newcolumntype{Y}{>{\raggedleft\arraybackslash}X}

\usepackage{threeparttable}
\usepackage{enumitem}

\usepackage{comment}

\biboptions{square,sort&compress,numbers}

\usepackage{graphicx}
\usepackage{floatrow}
\graphicspath{{graphics/}}

\usepackage{color}

\usepackage{booktabs}

\def\co{CO\textsubscript{2}}
\def\el{\textsubscript{el}}
\def\th{\textsubscript{th}}

\begin{document}

\begin{frontmatter}

  \title{PyPSA-ZA: Investment and operation co-optimization of integrating wind and solar in South Africa at high spatial and temporal detail}

\author[fias]{Jonas~Hörsch\corref{corrauthor}}
\cortext[mycorrespondingauthor]{Corresponding author}
\ead{hoersch@fias.uni-frankfurt.de}

\author[csir]{Joanne Calitz}
\ead{JRCalitz@csir.co.za}

\address[fias]{Frankfurt Institute for Advanced Studies, 60438~Frankfurt~am~Main, Germany}
\address[csir]{CSIR, Energy Centre, Council for Scientific and Industrial Research, Meiring Naude Road, Pretoria, South Africa}

\begin{abstract}
  South Africa is in the fortunate position of having access to high-quality wind and solar resources promising energy production costs well below the ones projected for future installations of thermal generation; yet the feasibility of a low carbon energy system is still under scrutiny. This study introduces a spatially and temporally resolved techno-economic model co-optimising generation and transmission capacity investments and operation. 27 supply regions in South Africa are equipped with a year of time-series in hourly resolution based on the IRP2016 base-case assumptions. Eight different scenarios representing different restrictions on renewable generation and mitigation efforts of GHG emissions are evaluated and discussed. 70\% of energy from renewable sources were found to be cost-optimally integrated using flexible gas generation and the expansion of several transmission corridors. A 95\% \co\ reduction leads to a moderate 20\% cost increase. The open-source nature of the model and restriction to freely available data encourages an accessible and transparent discussion about the future South African energy system, primarily based on renewable wind and solar resources.
\end{abstract}


\end{frontmatter}

\section{Introduction}

South Africa was attested high-quality wind and solar resources with potential for very high load factors and a weak seasonality that simplifies their integration into the electricity system~\cite{windpv-iwes}. Several GW of solar PV capacities can be integrated without major restructuring of the electricity grid~\cite{poeller2015,bofinger2016}. Nevertheless, the latest Integrated Resource Plan by the South African Department of Energy (DOE) proposes only a slow uptake of renewable energy sources and bases its electricity plan on new Nuclear and Coal power stations~\cite{doe-irp2016,csir-comments-irp2016}. Meanwhile, South Africa pledged to reduce its GHG emissions starting from at latest 2025, and the electricity sector is an easy target. This paper presents a spatially and temporally resolved model for performing least-cost techno-economic investment and operation optimisations of the South African electricity system. In 27 supply regions the hourly demand in the reference year 2040 has to be met by existing Coal, Nuclear or Hydro generation capacities or by generation from new open-cycle or closed-cycle gas turbines, coal plants, nuclear reactors, wind turbines or solar PV panels. The shape of the demand time-series as well as of renewable generation availability derive from historical load and weather data exhibiting the actual spatio-temporal correlations. The technology portfolio is complemented by battery storage and pumped hydro storage. The model is exlusively based on freely available data and open technologies.

In Section \ref{sec:methods-model} and \ref{sec:methods-data} the data sources and processing methods are presented; the optimisation results are presented in Section \ref{sec:results}; limitations of the model are discussed in Section \ref{sec:criticalappraisal}; conclusions are drawn in Section \ref{sec:conclusions}.

\section{Methods: Model}
\label{sec:methods-model}

This study models a future, renewable South African electricity network geared towards forecasts of the reference year 2040. The PyPSA framework underlying the PyPSA-ZA model implements a partial equilibrium model optimising both short-term operation and long-term investment in the energy system as a linear problem using the linear power flow equations~\cite{pypsa,pypsa-0.10}. It determines the optimal generation fleet to serve hourly electricity demand in the reference year. The intermediate build-up steps from today to 2040 are not assessed.

\subsection{Objective function}
The model minimises the total annual system costs as
\begin{equation}
  \min_{\substack{G_{n,s},F_\ell,\\ g_{n,s,t},f_{\ell,t}}} \left[ \sum_{n,s} c_{n,s} G_{n,s} + \sum_{\ell} c_{\ell} F_{\ell} + \sum_{n,s,t} o_{n,s} g_{n,s,t} \right],
\end{equation}
and consists of the capacities $G_{n,s}$ at each bus $n$ for generation and storage technologies $s$ and their associated annualized fixed costs $c_{n,s}$, the dispatch $g_{n,s,t}$ of the unit at time $t$ and the associated variable costs $o_{n,s}$. Further, the branch capacities $F_\ell$ for each branch $\ell$ and their annualized fixed costs $c_\ell$ are also included. The branch flows $f_{\ell,t}$ do not contribute to the total costs. The optimisation is run over all hours $t$ in a year with varying weather and demand conditions.

\subsection{Generator constraints}
The dispatch of conventional generators $g_{n,s,t}$ is constrained by their capacity $G_{n,s}$
\begin{equation}
  0 \leq g_{n,s,t} \leq G_{n,s} \qquad \forall\, n,s,t~.
\end{equation}
The dispatch is limited by ramp rate constraints $ru_{n,s}$ and $rd_{n,s}$ per unit of the generator nominal power\footnote{Currently only implemented for existing Coal power stations.},
\begin{equation}
   -rd_{n,s}\,G_{n,s} \leq (g_{n,s,t} - g_{n,s,t-1}) \leq ru_{n,s}\,G_{n,s}~.\label{eq:ramp}
\end{equation}

The maximum producible power of renewable generators depends on the weather conditions, which is expressed as an availability $\bar{g}_{n,s,t}$ per unit of its capacity:
\begin{equation}
 0 \leq  g_{n,s,t} \leq \bar{g}_{n,s,t} G_{n,s} \qquad \forall\, n,s,t~.
\end{equation}

The power capacity $G_{n,s}$ is also subject to optimization up to a maximum installable potential $\bar{G}_{n,s}$ restricted by available geographic area:
\begin{equation}
 0 \leq G_{n,s}\leq \bar{G}_{n,s} \qquad \forall\, n,s~.
\end{equation}

\subsection{Storage operation}

The energy levels $e_{n,s,t}$ of all storage units have to be consistent between all hours and are limited by the storage energy capacity $E_{n,s}$
\begin{align}
  e_{n,s,t} &= \eta_{n,s,0} e_{n,s,t-1} + \eta_{n,s,+} \left[g_{n,s,t}\right]^+  -  \eta_{n,s,-}^{-1} \left[g_{n,s,t}\right]^- + g_{n,s,t,\textrm{inflow}} - g_{n,s,t,\textrm{spillage}}~, \\
       0 &\leq  e_{n,s,t} \leq E_{n,s}\qquad \forall\, n,s,t~.
\end{align}
Positive and negative parts of a value are denoted as $[\cdot]^+ = \max(\cdot,0)$, $[\cdot]^{-} = -\min(\cdot,0)$. The storage units have a charging efficiency $\eta_{n,s,+}$, a discharging efficiency $\eta_{n,s,-}$, in-flow (e.g. river in-flow in a reservoir) and spillage. The energy level is set to be cyclic, i.e. $e_{n,s,t=0} = e_{n,s,t=T}$.

\subsection{Power balance and transmission constraints}
\label{sec:powerbalance}

The (inelastic) electricity demand $d_{n,t}$ at each bus $n$ must be met at each time $t$ by either local generators and storage or by the flow $f_{\ell,t}$ from a transmission line $\ell$
\begin{equation}
  \sum_{s} g_{n,s,t} - d_{n,t} = \sum_{\ell} K_{n\ell}\, f_{\ell,t}  \quad \leftrightarrow \quad \lambda_{n,t} \qquad \forall\, n,t~, \label{eq:balance}
\end{equation}
where $K_{n\ell}$ is the incidence matrix of the network and $\lambda_{n,t}$ is the marginal price at the bus. This equation implements Kirchhoff's Current Law (KCL). For the physicality of network flows, additionally Kirchhoff's Voltage Law (KVL)
\begin{equation}
  \sum_{\ell} C_{\ell c} x_{\ell} f_{\ell,t} = 0 \qquad \forall c,t~, \label{eq:kvl}
\end{equation}
expressed with a cycle basis $C_{\ell c}$ and the series inductive reactance $x_\ell$ must be enforced~\cite{2017arXiv170401881H}.

The flows in all branches are constrained by their capacities $F_{\ell}$
\begin{equation}
  |f_{\ell,t}| \leq F_{\ell} \qquad \forall\,\ell,t~.
\end{equation}

Since the expansion of line capacities $F_\ell$ representing the addition of new circuits leads to decreasing line impedances $x_\ell$, Equation~\ref{eq:kvl} introduces in principle a bilinear coupling. We maintain the linearity and therefore computational speed by solving the optimisation problem with fixed impedances $x_\ell$, updating them and re-solving in up to \num{10} iterations to ensure convergence, following the methodology of \cite{Hagspiel}.

\subsection{\co\ emission constraint}
\label{sec:co2}

\co\ emissions in the \emph{\co\ limit} scenarios are limited by a cap $\textrm{CAP}_{CO2} = 10\,\mathrm{MtCO_2}$ corresponding to a 95\% reduction from today's emissions by the electricity sector, implemented using the specific emissions $e_{s}$ in \co-tonne-per-MWh of the fuel $s$ and the efficiency $\eta_{n,s}$ of the generator:
\begin{equation}
  \sum_{n,s,t} \frac{1}{\eta_{n,s}} g_{n,s,t}\cdot e_{s} \leq  \textrm{CAP}_{CO2} \quad \leftrightarrow \quad \mu_{CO2}~. \label{eq:co2cap}
\end{equation}
$\mu_{CO2}$ is the shadow price of this constraint.

\section{Methods: Data}
\label{sec:methods-data}

\subsection{Network topology}

\begin{figure}
  \includegraphics[width=8cm]{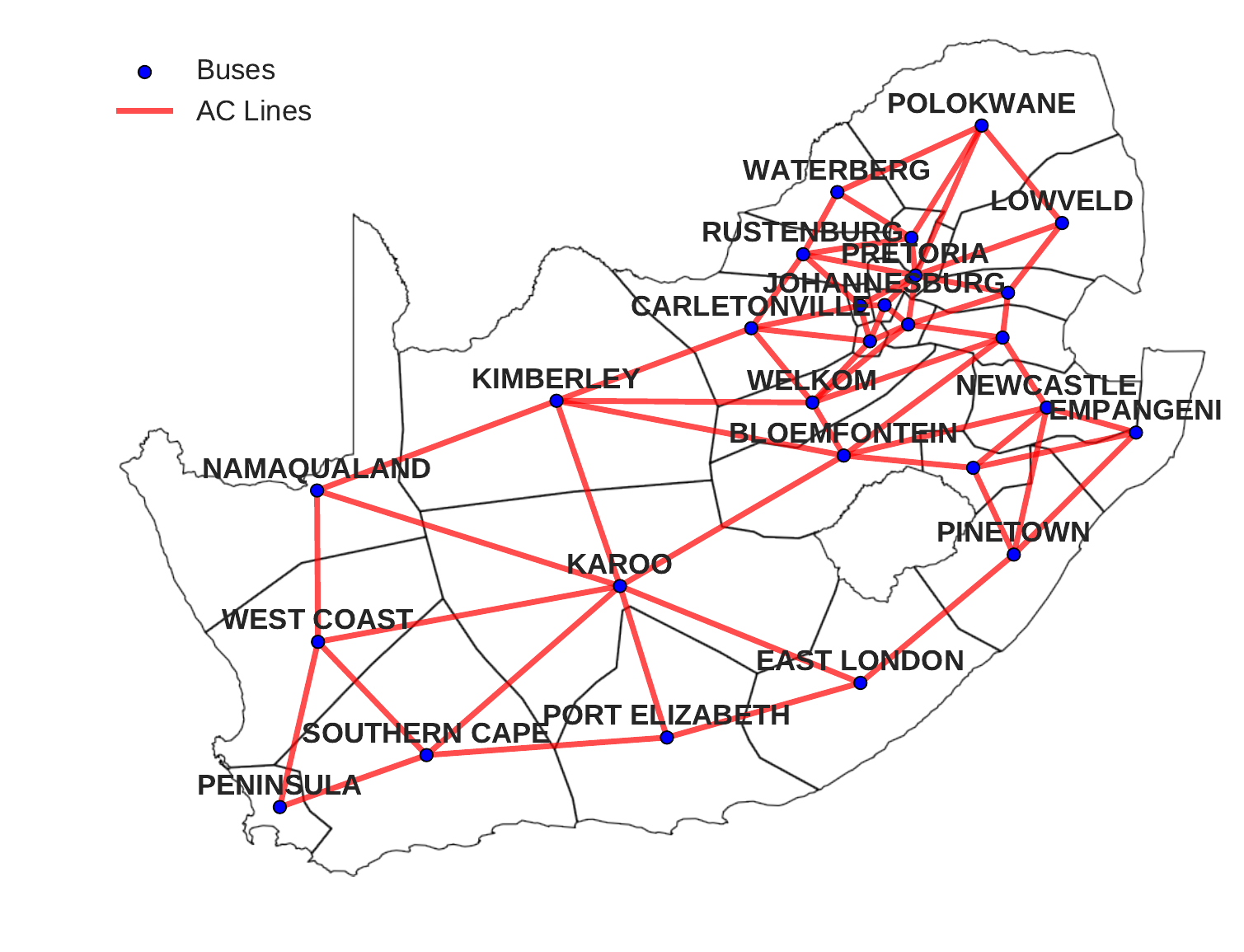}
  \caption{Network model. 27 supply areas as buses. Adjacent supply areas have
    been connected with typical $380\,\mathrm{kV}$-lines.}
  \label{fig:network-model}
\end{figure}

The South African transmission grid has been divided into \num{27} supply areas. They abstract underlying network topology to deliver electricity to between \num{0.1} and \num{7} million people each. The capacities necessary to connect them strongly depend on the future layout of generation capacities in South Africa and is ideally optimised jointly. The simplified network model shown in Figure~\ref{fig:network-model} consists of one bus in each supply area located at the centroid of the geographical area\footnote{The buses for the supply areas NAMAQUALAND and PRETORIA have been moved slightly away from the centroid back into the area.} and transmission lines connecting all adjacent areas. The lines are modelled as standard overhead AC lines with length of \num{1.25} times geodesic distance and the electrical parameters of a typical $380~\mathrm{kV}$-line~\cite{oeding2011-book}. The current inter-area transmission capacities have been determined by summing the capacities of all transmission lines above and including $275\,\mathrm{kV}$ of the Electricity Transmission and Distribution Grid Map (2017) dataset by the World Bank Group~\cite{africagrid}. The model may add line capacities between all directly adjacent supply regions.

\subsection{Electricity demand}
The shape of the electricity demand as an hourly aggregated time-series for the four years 2009-2013 was provided by ESKOM \cite{eskom2016-system-energy} and was scaled linearly to the total yearly electricity demand forecast of \SI{428}{TWh} in the high demand base case of the Integrated Resource Plan 2016 (IRP2016) by the Department of Energy~\cite{doe-irp2016}. The year 2012 was extracted from the time-series and distributed to the 27 buses in proportion to population in each supply area, as determined by aggregation of the South African population density published by World POP \cite{afripop}.

\subsection{Existing power stations} 
The currently existing power stations in South Africa have been compiled from Eskom holdings~\cite{eskom-holdings} listed online and the IRP2016~\cite{doe-irp2016}. It is expected that of these $51\,\mathrm{GW}$ generation capacities by the reference year 2040 about half will have decommissioned save for $19.9\,\mathrm{GW}$ of coal stations, $0.7\,\mathrm{GW}$ of hydro reservoir stations, $2.9\,\mathrm{GW}$ of pumped hydro storage and the Koeberg Nuclear power station with $1.86\,\mathrm{GW}$. The remaining power plants are attached at the bus of their supply area with marginal and capital costs taken from their fixed and variable operation and maintenance costs plus fuel costs. Additionally, the Cahora Bassa reservoir station in Mozambique which has an installed capacity of $2.075\,\mathrm{GW}$ is attached as a $1.5\,\mathrm{GW}$ hydro reservoir station at the POLOKWANE supply area where the two HVDC lines from the dam end in South Africa.

Pumped storage units as well as hydro reservoirs receive a weather dependent hydro-electric in-flow. The re-analysis weather dataset CFSv2 by the US National Oceanic Atmospheric Administration~\cite{saha} provides water run-off on a 0.2 x 0.2 deg. spatial raster ($x \in \mathcal{X}$) in hourly resolution since 2011. Following the methodology in~\cite{kies2016,Schlachtberger2017}, we aggregated the total potential energy at height $h_x$ of the run-off data $\mathcal{R}_x$ in South Africa as well as in Mozambique $c$ as a simple proxy to their hydro-electric in-flow time-series
\begin{equation}
  G^H_{c}(t) = \mathcal{N} \sum_{x \in \mathcal{X}(c)} h_x \mathcal{R}_x(t)
\end{equation}
where $\mathcal{N}$ is chosen so that $\int_{t} G^H_{c}(t) \,\mathrm{d}t$ matches the EIA annual hydroelectricity generation~\cite{EIA-hydro-netgen-za-mz}. The in-flow is distributed to all run-of-river and reservoir capacities in proportion to their power capacity.

\subsection{Expandable power generation}

The existing generation and storage capacity at full availability amounts to $26.8\,\mathrm{GW}$, barely half the peak load of $62.3\,\mathrm{GW}$; opening the possibility to a large-scale restructuring of the generation technology mix. The model can build wind turbines and solar panels within restrictive landuse limits and open cycle and closed cycle gas turbines (OCGT/CCGT), nuclear reactors and coal power stations at suitable locations and battery storage units at every bus.

\subsubsection{Wind and solar PV generation}
\label{sec:windpv}

The per-unit availability time-series $\bar{g}_{n,s,t}$ for wind and solar photovoltaics for each supply region in the year 2012 have been calculated within the Wind and Solar PV Resource Aggregation Study for South Africa \cite{windpv-iwes}. Re-analysis wind speeds of the Wind Atlas for South Africa (WASA)~\cite{wasa2015} were converted to available wind power feed-in for a capacity layout based on a selection of five representative turbines suitable for low to high wind speeds. Solar PV feed-in has been generated using satellite-imaging based global irradiation with the SODA model~\cite{soda2006} for fixed tilted PV installations.

For the estimation of maximal installable wind and solar PV generation capacities $\bar{G}_{n,s}$, we determine the available area $A_n$ in each supply region by restricting to landuse types ``Grassland'', ``Low Shrubland'' and ``Bare none vegetated'' in the South African National Land-Cover Dataset~\cite{landcover} that are not part of protected or conservation areas in~\cite{sapad,sacad}. 80\% of this area is split equally between wind and solar PV installations, leading to
\begin{equation}
  \bar{G}_{n,w} = 0.5 \cdot 0.8 \cdot 10\,\mathrm{MW/km^2} \cdot A_{n}~, \qquad \bar{G}_{n,s} = 0.5 \cdot 0.8 \cdot 33\,\mathrm{MW/km^2} \cdot A_{n}~,
\end{equation}
where $10\,\mathrm{MW/km^2}$ and $33\,\mathrm{MW/km^2}$ are the technical potential densities of wind and solar PV respectively~\cite{windpv-iwes}. The factor $0.8$ was chosen so that these simplified assumptions applied to the renewable energy development zones (REDZ)~\cite{redz} alone lead approximately to the same available area of $53.4 \cdot 10^3\,\mathrm{km^2}$ reported by the detailed evaluation of the resource aggregation study.

Since the resulting renewable potentials overshoot the demand requirements by an order of magnitude, we evaluate in the following two different scenarios, in the \emph{REDZ} scenario renewable expansion is only allowed in the REDZ areas~\cite{redz}, and a largely unrestricted scenario \emph{CORRIDORS} where the area is confined along the main transmission corridors defined by \cite{corridors}.

\subsubsection{Storage}
In addition to the pumped hydro storage installed today, the model may build battery storage units at every bus. Their charging and discharging efficiencies, as well as cost assumptions for their power and energy storage capacities are taken from~\cite{inputs-technology-costs}. It is assumed that the charging and discharging power capacities of a unit are equal, and the energy capacity $E_{n,s} = \bar{h}_s \cdot G_{n,s}$ is proportional to this power capacity. The factor $\bar{h}_s$ determines the time for charging or discharging the storage completely at maximum power, and is set to $\bar{h}_s = 3\,\mathrm{h}$.

\subsubsection{Non-renewable generation}
\label{sec:non-renewable-gen}

The model contains four different types of expandable fossil fuel generators: OCGT and CCGT can be build at Saldana Bay, Richards Bay and Coega, nuclear reactors are installable at Thyspunt and Koeberg and new coal capacities can be added in Waterberg. In the \emph{\co\ limit} scenarios their usage is subject to the \co\ cap of Equation~\ref{eq:co2cap}, while in the \emph{Emission prices} scenarios their marginal costs include externality costs for specific emissions. The \emph{business-as-usual (BAU)} scenario enforces the construction of at least \SI{10}{GW} of nuclear and coal capacities, each, in line with current energy plans.

\subsection{Cost assumptions}
\label{sec:costs}
Investment, fixed and variable operation and maintenance (FOM and VOM) costs and fuel prices and efficiencies for all assets are listed in Table~\ref{tab:costs}. They mirror the IRP2016 draft, except for the overnight capacity costs for the renewable technologies and battery storage \cite{doe-irp2016}. The CSIR calculated the capacity costs for Wind and Solar by de-annualizing the winning bids in the Renewable Independent Power Producer Procurement Programme (REIPPPP) in 2015 as a dependable proxy to today's installation costs to \SI{13,250}{R/kW} for onshore wind and \SI{9,243}{R/kW} for fixed photovoltaics~\cite{inputs-technology-costs}. BNEF forecasts a reduction of the levelized cost of electricity of PV by $2/3$, by $1/2$ for onshore wind and by $3/4$ for lithium-ion batteries~in respect to the reference year $2040$ \cite{bnef-neo2017}. As a conservative estimate, the proposed capacity costs in this paper reflect half of this reduction potential, while discussing the full projected interval at the end of Section~\ref{sec:results}. For the annualisation of overnight costs a discount rate of 8\% is used. The transmission investment of a line with length $l_{\ell}$ is $\SI{6000}{R/MWkm} \cdot l_\ell f_{n-1}$ with an $n-1$ security factor $f_{n-1} = 1/0.7$.

In the \emph{Emission prices} and \emph{\co\ limit} scenarios the minimized objective additionally includes the externality costs for specific emissions as referenced in the Integrated Energy Plan by the DOE: \SI{0.27}{R/kgCO_2}, \SI{7.6}{R/kgSO_2}, \SI{4.5}{R/kgNO_x}, \SI{41.5}{R/tHg} and \SI{11.3}{R/kgPM} (particulate matter)~\cite{doe-iep-report2016}. Since the prices are dual to emission caps, in the \emph{\co\ limit} scenarios the \co\ price is omitted from the optimization.

\begin{table}
\centering
\caption{Capital and marginal costs of extendable components} \label{tab:costs}
\resizebox{\linewidth}{!}{%
\begin{tabular}{@{}lrlrrrrr@{}}
  \toprule
  Quantity  &  \multicolumn{4}{c}{Capital costs} & \multicolumn{3}{c}{Marginal costs} \\
                  &Overnight   &Unit & FOM    & Lifet. & VOM       & Fuel cost & Effic. \\
                          & cost {\footnotesize [R]}   &     & {\footnotesize [\%/a]} & {\footnotesize [a]}      & {\footnotesize [R/MWh\el]} & {\footnotesize [R/MWh\th]} &  \\
  \midrule
  Wind            &10000    &kW\el  & 4   & 20 & & &  \\
  Solar PV        &6000    &kW\el  & 2   & 25 & & & \\
  OCGT            &8173    &kW\el  & 2   & 30 & 2.4 & 540 & 0.31  \\
  CCGT            &8975    &kW\el  & 1.8 & 30 & 22  & 540 & 0.49  \\
  Coal            &35463   &kW\el  & 2.6 & 30 & 80  & 98  & 0.37  \\
  Nuclear         &60447   &kW\el  & 1.6 & 60 & 37  & 29  & 0.34  \\
  Battery stor.   &12000    &kW\el  & 2.5 & 20 & 3.2 &    & 0.89 \\
  Transm. lin.    &6000    &MWkm   & 2   & 40 & & &\\
  \bottomrule 
\end{tabular}
}
\end{table}


\section{Model results}
\label{sec:results}

The model is optimized for eight different scenarios. The base case at market
costs, two different emission mitigation scenarios by emission prices and 95\%
\co\ reduction cap and a business-as-usual scenario with forced coal and nuclear. For each of these, we compare one scenario with renewable generators constrained to \emph{REDZ} and another one with the whole transmission \emph{CORRIDORS} available.

\begin{figure}[p]
\includegraphics[width=\linewidth]{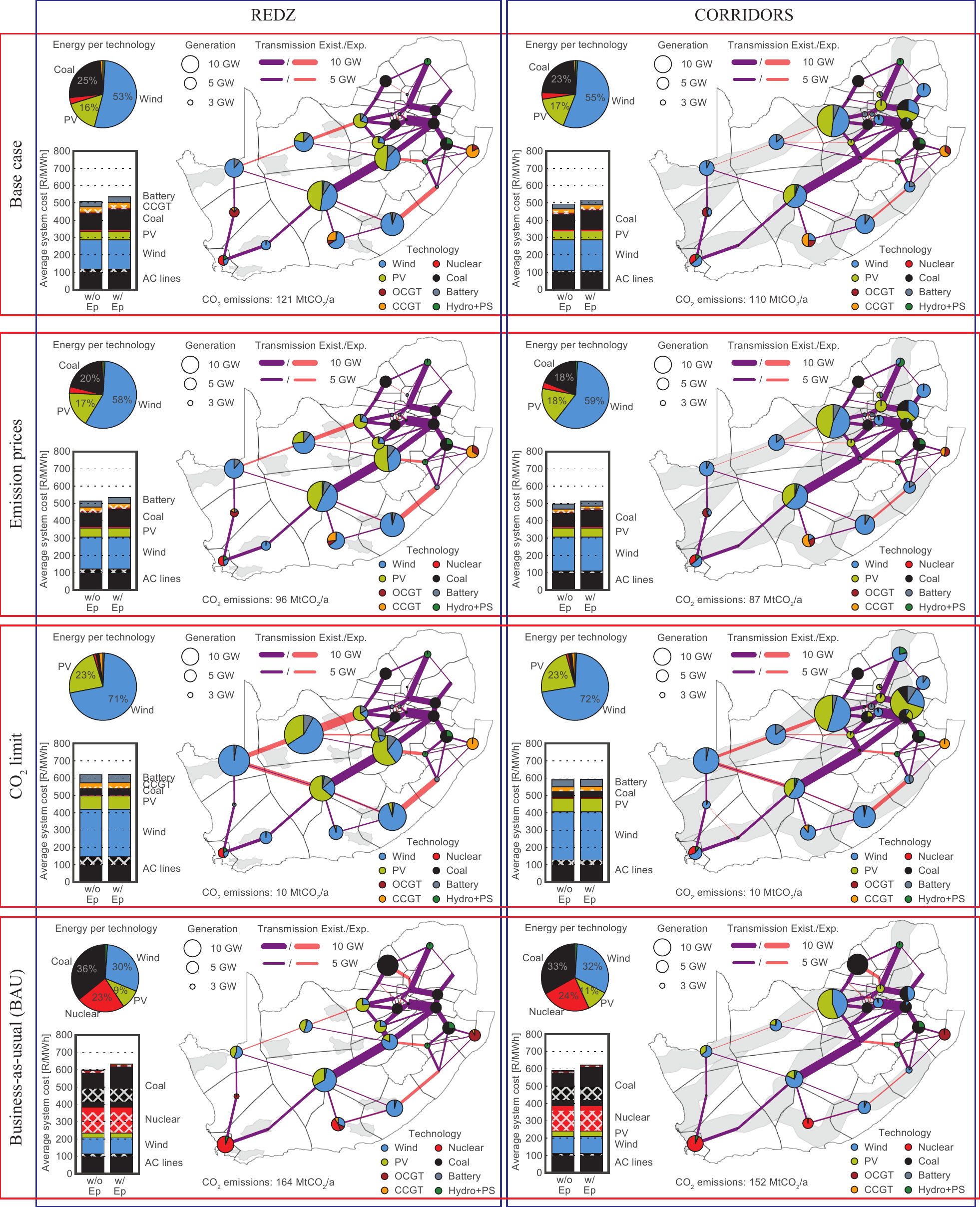}
\caption{Matrix of scenario results. For each scenario, the top-left pie-chart shows the mix of generated energy, the bar-charts decompose the cost per MWh into the technologies, the map shows the distribution of generation, storage and existing and expanded transmission capacities. The bar-charts distinguish costs with and without externality costs by emissions and separate the cost for new transmission and conventional generation capacities with white xx hatches. The light gray map background highlights the areas available for renewable installations (before landuse restrictions) in the \emph{REDZ} and \emph{CORRIDORS} scenarios (see Sec.~\ref{sec:windpv}). The \emph{Base case} at the top optimizes at market costs, while the \emph{Emission prices} and \emph{\co\ limit} scenarios include externality costs by emissions (see Sec.~\ref{sec:costs}). In the \emph{\co\ limit} scenarios there is a strict $10\,\mathrm{MtCO_2}$ cap (see Sec.~\ref{sec:co2}). The \emph{BAU} scenario enforces \SI{10}{GW} of nuclear and coal generation capacities (see Sec.~\ref{sec:non-renewable-gen}).}
\label{fig:results}
\end{figure}

\begin{figure}
\includegraphics[width=\linewidth]{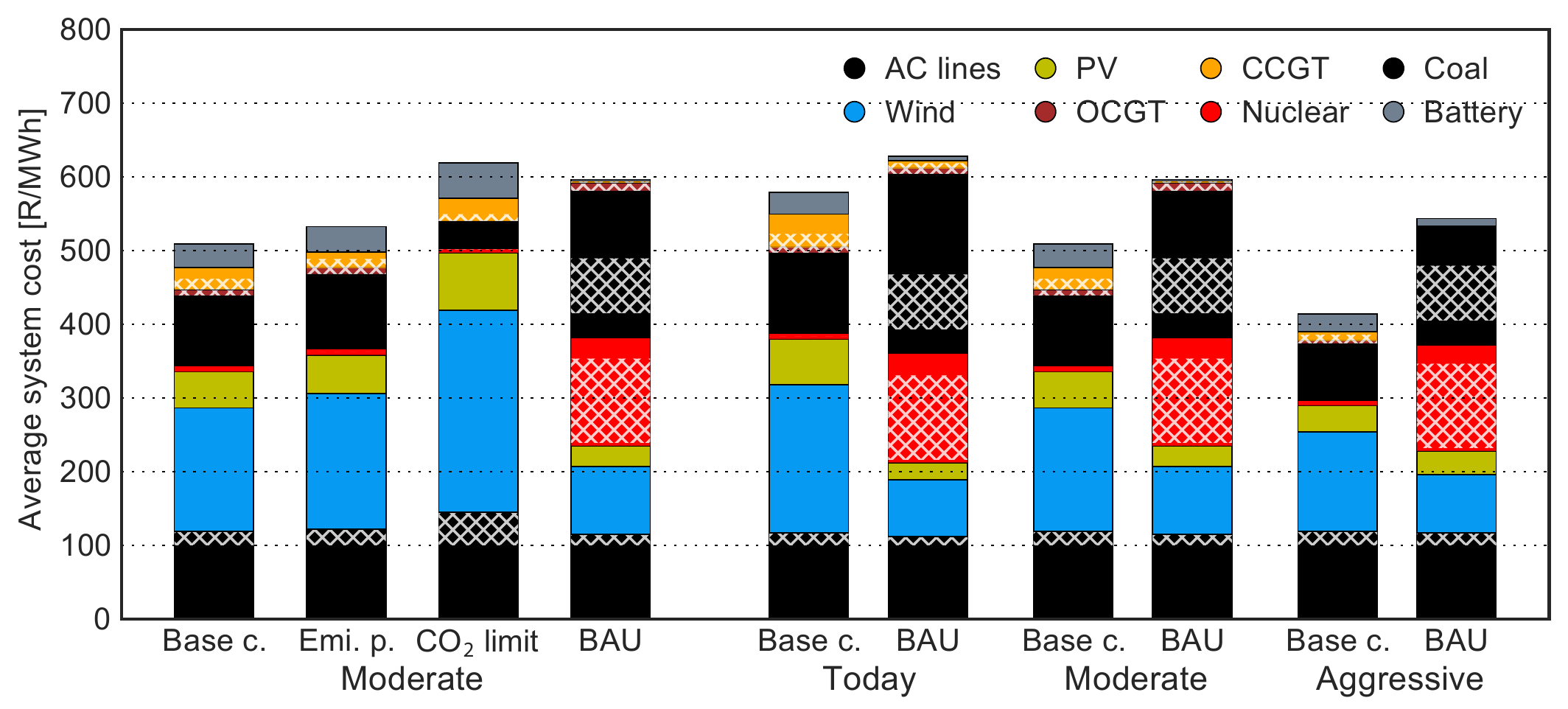}
\caption{Cost comparison of scenarios with wind and PV capacities confined to \emph{REDZ}. The four scenarios on the left are different emission mitigation strategies under a moderate cost reduction, the six scenarios on the right show the \emph{base case} and the \emph{business-as-usual} case for three different cost assumptions (see Sec.~\ref{sec:costs}). The bar-charts decompose the average system cost into technologies. New transmission and new conventional generation capacities are highlighted by white xx hatches.}
\label{fig:costs-overview}
\end{figure}

The energy mix of the least-cost solutions of the \emph{base case} shown at the top of Figure~\ref{fig:results} is already strongly based on renewable sources with 53\% wind and 16\% solar PV. Their varying in-feed is balanced by \SI{11}{GW} new battery storage units and \SI{12}{GW} of flexible gas turbines, supplying a mere 1.4\% of demand. If the renewable generators can also be distributed to the eastern part of South Africa (within the \emph{CORRIDORS}), the additional geographical smoothing by long distances allows to integrate 2 percentage points more wind energy with only \SI{10}{GW} batteries and \SI{10}{GW} OCGT. The cost per MWh lie between $520\,\mathrm{R}$ and $500\,\mathrm{R}$. The externality costs of the IEP2016 introduced in Section~\ref{sec:costs} add \SI{25}{R/MWh}; of which \SI{3}{R/MWh} can be saved with an optimised capacity layout by supplying 5 percentage points of demand generated in coal power stations instead with energy from wind turbines, leading to a 20\% reduction of \co\ from the base case or to approximately 50\% compared to today's emission levels.


A deep decarbonisation to a 95\% \co\ reduction increases the model costs by about 20\%. The shadow price $\mu_{CO2}$ to the \co\ emission constraint from Equation~\eqref{eq:co2cap} captures the linear sensitivity of the total system cost to the \co-cap. Its value $1800-1870\,\mathrm{R/tCO_2}$ gives an indication of the \co-price equivalent to the cap, which is about 7 times the value assumed in IEP2016.

The main cost driver in the \emph{\co\ limit} scenarios is the provision of flexibility for integrating 94\% of varying renewable generation: In the \emph{REDZ} scenario additional transmission capacities and battery storage units account for a cost increase by $43\,\mathrm{R/MWh}$, while the remaining $67~\mathrm{R/MWh}$ are due to a renewable capacity over-installation with about 19\% of curtailment. This hints at a great potential for coupling to the other energy sectors, and indeed preliminary results from including a passenger transport sector based on battery electric vehicles suggest that their batteries can replace the dedicated battery storage units entirely.

Lifting the confinement of renewable generation from \emph{REDZ} to \emph{CORRIDORS} reduces the costs for batteries and AC lines by about $15~\mathrm{R/MWh}$ each. The need for flexibility reduces and generation is brought closer to load centres around Johannesburg requiring less transmission expansion.

No new coal or nuclear power stations are chosen in any of the previous scenarios; to assess the extra cost incurred by the current electricity plans, we force the build-out of at least \SI{10}{GW} new coal and nuclear capacities in the business-as-usual scenarios, increasing the cost to \SI{600}{R/MWh} or even \SI{630}{R/MWh}, if emission prices are taken into account. In the latter case, it is more expensive than the \emph{\co\ limit} scenario, while leading to 15-16 times as much \co\ emissions.

The hotspots for transmission expansion are relatively stable across all scenarios, the most important lines are the new connection between EAST LONDON and PINETOWN in the South East ($10-14\,\mathrm{GW}$ in the \emph{\co\ limit} scenarios) and lines from NAMAQUALAND and KIMBERLEY in the North-West to the well-connected load-intensive central eastern South Africa around Johannesburg, also making use of the current grid backbone to PENINSULA passing through KAROO.

Fig.~\ref{fig:costs-overview} brings the cost compositions of the \emph{REDZ} scenarios already shown in Fig.~\ref{fig:results} side-by-side for easier comparison and additionally presents the influence of different cost developments on the \emph{base case}: While the average system cost in 2040 is still uncertain, the optimal composition is equally defined by high levels of renewable generation with gas turbines and battery storage units for flexibility without additional nuclear or coal power stations.

\section{Critical Appraisal}
\label{sec:criticalappraisal}

There are a number of simplifications in the modelling presented here that may affect the conclusions.

Available data on the electricity demand time series is not spatially disaggregated; assuming, as we have done, that the load time series shape and demand growth is the same at each node ignores local differences. Information on the storage capacity of existing hydro reservoir dams was not available and assumed to be \SI{20.6}{hrs} like the average pumped hydro storage capacity. Similarly, hydro-electric in-flow time series are based on country-wide approximations, ignoring local topography and basin drainage; in principle a full hydrological model should be used.

The network is represented by standard transmission lines between 27 zones in a single voltage layer. Grid upgrades to prevent intrazonal congestion in the transmission and distribution networks and between them can not be assessed within the model.

Only a single year (2012) has been modelled, so the model may be overfitting to characteristics of this particular year.

Additional aspects, such as reserve power, stability, transmission losses, efficiency savings, demand side management and sector-coupling have not been considered.

These issues will be addressed in forthcoming studies.

\section{Conclusions}
\label{sec:conclusions}

In this paper the techno-economic model PyPSA-ZA has been presented that optimises investment and operation costs of the South African electricity system with 27 buses in hourly resolution. Its focus lies on deep decarbonisation scenarios.

By treating operations based on the linear power flow endogeneously, a significant cost reduction potential from improved coordination within the electricity system is uncovered, especially to the benefit of varying renewable energy sources. Solutions integrating a 70\% share of renewables were found to be cost-optimal with flexible gas generation, battery storage units and by developing new transmission corridors between EAST LONDON and PINETOWN and from the resource-rich North-western supply regions. A 95\% \co\ reduction relative to today's emission levels is achievable within the model assumptions by a moderate 20\% increase in cost. Enforcing the build-out of coal and nuclear capacities like current energy plans leads to 15 times higher \co\ emissions at similar costs.

The model builds exclusively on freely available data and open technologies making it suitable for an iterative and collaborative improvement by many stakeholders. We hope that it will contribute towards a transparent discussion of the future needs of the South African and Southern African energy system.

\section*{Acknowledgements}

We thank Tom Brown, Jarrad Wright, Tobias Bischof-Niemz and Crescent Mushwana for support and helpful discussions. This research was conducted as part of the CoNDyNet project, which is supported by the German Federal Ministry of Education and Research under grant no. 03SF0472C. The responsibility for the contents lies solely with the authors.

\bibliography{pypsa-za}

\begin{thebibliography}{10}
\expandafter\ifx\csname url\endcsname\relax
  \def\url#1{\texttt{#1}}\fi
\expandafter\ifx\csname urlprefix\endcsname\relax\def\urlprefix{URL }\fi
\expandafter\ifx\csname href\endcsname\relax
  \def\href#1#2{#2} \def\path#1{#1}\fi

\bibitem{windpv-iwes}
D.~S. Bofinger, B.~Zimmermann, A.-K. Gerlach, D.~T. Bischof-Niemz, C.~Mushwana,
  \href{https://www.csir.co.za/csir-energy-centre-documents}{{Wind and Solar PV
  Resource Aggregation Study for South Africa}}, available online. (2016).
\newline\urlprefix\url{https://www.csir.co.za/csir-energy-centre-documents}

\bibitem{poeller2015}
M.~P\"oller, M.~Obert, M.~Geeen, Analysis of options for the future allocation
  of {PV} farms in {S}outh {A}frica (2015).

\bibitem{bofinger2016}
S.~Bofinger, K.~Knorr, B.~Zimmermann,
  \href{https://www.csir.co.za/csir-energy-centre-documents}{{Analysis of the
  curtailments needed to integrate wind and solar PV power into substations of
  the South African transmission grid}}, available online. (2016).
\newline\urlprefix\url{https://www.csir.co.za/csir-energy-centre-documents}

\bibitem{doe-irp2016}
{Republic of South Africa Department of Energy},
  \href{http://www.energy.gov.za/IRP/2016/Draft-IRP-2016-Assumptions-Base-Case-and-Observations-Revision1.pdf}{Integrated
  resource plan update: Assumptions, base case results and observations},
  available online (accessed August 2017). (2016).
\newline\urlprefix\url{http://www.energy.gov.za/IRP/2016/Draft-IRP-2016-Assumptions-Base-Case-and-Observations-Revision1.pdf}

\bibitem{csir-comments-irp2016}
{CSIR Energy Centre},
  \href{https://www.csir.co.za/csir-energy-centre-documents}{{Formal comments
  on the South African Integrated Resource Plan (IRP) Update Assumptions, Base
  Case and Obserations 2016}}, available online (accessed August 2017). (2017).
\newline\urlprefix\url{https://www.csir.co.za/csir-energy-centre-documents}

\bibitem{pypsa}
T.~{Brown}, J.~{Hörsch}, D.~{Schlachtberger}, {PyPSA: Python for Power System
  Analysis}, submitted. (Jul 2017).
\newblock \href {http://arxiv.org/abs/1707.09913v2}
  {\path{arXiv:1707.09913v2}}.

\bibitem{pypsa-0.10}
T.~Brown, J.~H\"orsch, D.~Schlachtberger, \href{http://pypsa.org}{{PyPSA:
  Python for Power System Analysis version 0.10.0}} (Apr 2017).
\newblock \href {http://dx.doi.org/10.5281/zenodo.582307}
  {\path{doi:10.5281/zenodo.582307}}.
\newline\urlprefix\url{http://pypsa.org}

\bibitem{2017arXiv170401881H}
J.~{H{\"o}rsch}, H.~{Ronellenfitsch}, D.~{Witthaut}, T.~{Brown}, {Linear
  Optimal Power Flow Using Cycle Flows}, ArXiv e-prints.
\newblock \href {http://arxiv.org/abs/1704.01881} {\path{arXiv:1704.01881}}.

\bibitem{Hagspiel}
{Hagspiel, S.}, {J\"agemann, C.}, {Lindenburger, D.}, {Brown, T.},
  {Cherevatskiy, S.}, {Tr\"oster, E.}, Cost-optimal power system extension
  under flow-based market coupling, Energy 66 (2014) 654--666.

\bibitem{oeding2011-book}
D.~Oeding, B.~Oswald, Elektrische Kraftwerke und Netze, 7th Edition, Springer,
  2011.

\bibitem{africagrid}
C.~Arderne, \href{http://africagrid.energydata.info/}{{Africa - Electricity
  Transmission and Distribution Grid Map}}, available online (accessed August
  2017). (Mar. 2017).
\newline\urlprefix\url{http://africagrid.energydata.info/}

\bibitem{eskom2016-system-energy}
Eskom, {System Energy 2009-13 Hourly}, available on request.

\bibitem{afripop}
{WorldPop}, {South Africa 100m Population} (2013).
\newblock \href {http://dx.doi.org/10.5258/soton/wp00246}
  {\path{doi:10.5258/soton/wp00246}}.

\bibitem{eskom-holdings}
\href{https://www.eskom.co.za/}{{Eskom Holdings}}, available online (accessed
  January 2017).
\newline\urlprefix\url{https://www.eskom.co.za/}

\bibitem{saha}
S.~et~al., The {NCEP} {C}limate {F}orecast {S}ystem {R}eanalysis, Bull. Amer.
  Meteor. Soc. 91~(8) (2010) 1015--1057.

\bibitem{kies2016}
{Kies, A.}, {Chattopadhyay, K.}, {von Bremen, L.}, {Lorenz, E.}, {Heinemann,
  D.}, {RESTORE 2050 Work Package Report D12: Simulation of renewable feed-in
  for power system studies.}, Tech. rep., RESTORE 2050, in preparation (2016).

\bibitem{Schlachtberger2017}
D.~Schlachtberger, T.~Brown, S.~Schramm, M.~Greiner, The benefits of
  cooperation in a highly renewable european electricity network, Energy.

\bibitem{EIA-hydro-netgen-za-mz}
{U.S. Energy Information Administration},
  \href{http://tinyurl.com/EIA-hydro-gen-ZA-MZ-2011-2014}{{Hydroelectricity Net
  Generation ZA and MZ 2011-2014}} (2017).
\newline\urlprefix\url{http://tinyurl.com/EIA-hydro-gen-ZA-MZ-2011-2014}

\bibitem{wasa2015}
A.~N.~H. et~al,
  \href{http://orbit.dtu.dk/files/107110172/DTU_Wind_Energy_E_0050.pdf}{{Mesoscale
  modeling for the Wind Atlas of South Africa (WASA) project}} (Feb 2015).
\newline\urlprefix\url{http://orbit.dtu.dk/files/107110172/DTU_Wind_Energy_E_0050.pdf}

\bibitem{soda2006}
B.~Gschwind, L.~Ménard, M.~Albuisson, L.~Wald, {Converting a Successful
  Research Project into a Sustainable Service: The Case of the SoDa Web
  Service} 21 (2006) 1555--1561.

\bibitem{landcover}
{GEOTERRAIMAGE (South Africa)},
  \href{https://egis.environment.gov.za/data_egis/node/109}{{2013-14 South
  African National Land-Cover Dataset}}, available online (accessed August
  2017). (2017).
\newline\urlprefix\url{https://egis.environment.gov.za/data_egis/node/109}

\bibitem{sapad}
{Republic of South Africa Department of Environmental Affairs},
  \href{https://egis.environment.gov.za/data_egis/node/109}{{SA Protected Areas
  Database (SAPAD\_OR\_2017\_Q2)}}, available online (accessed August 2017).
  (Jun 2017).
\newline\urlprefix\url{https://egis.environment.gov.za/data_egis/node/109}

\bibitem{sacad}
{Republic of South Africa Department of Environmental Affairs},
  \href{https://egis.environment.gov.za/}{{South Africa Conservation Areas
  Database (SACAD\_OR\_2017\_Q2)}}, available online (accessed August 2017).
  (Jun 2017).
\newline\urlprefix\url{https://egis.environment.gov.za/}

\bibitem{redz}
{Republic of South Africa Department of Environmental Affairs},
  \href{https://egis.environment.gov.za/}{{Wind and Solar PV Energy Strategic
  Environmental Assessment REDZ Database}}, available online (accessed August
  2017). (Mar 2017).
\newline\urlprefix\url{https://egis.environment.gov.za/}

\bibitem{corridors}
{Republic of South Africa Department of Environmental Affairs},
  \href{https://egis.environment.gov.za/}{{REDZs Strategic Transmission
  Corridors}}, available online (accessed August 2017). (Apr 2017).
\newline\urlprefix\url{https://egis.environment.gov.za/}

\bibitem{inputs-technology-costs}
\href{https://www.csir.co.za/documents/irp2016-inputstechnology-costsxlsx}{{IRP2016\_Inputs\_Technology-Costs
  (PUBLISHED)}}, available online. (2017).
\newline\urlprefix\url{https://www.csir.co.za/documents/irp2016-inputstechnology-costsxlsx}

\bibitem{bnef-neo2017}
{Bloomberg New Energy Finance},
  \href{https://about.bnef.com/new-energy-outlook/}{{New Energy Outlook 2017}},
  Tech. rep., executive summary freely available online. (2017).
\newline\urlprefix\url{https://about.bnef.com/new-energy-outlook/}

\bibitem{doe-iep-report2016}
{Republic of South Africa Department of Energy},
  \href{http://www.energy.gov.za/files/IEP/2016/Integrated-Energy-Plan-Report.pdf}{{Integrated
  Energy Plan Report}}, available online (accessed September 2017). (2016).
\newline\urlprefix\url{http://www.energy.gov.za/files/IEP/2016/Integrated-Energy-Plan-Report.pdf}

\end{thebibliography}

\end{document}